\documentclass[aps,pra,reprint]{revtex4-1}

\usepackage{amsmath,amsfonts,amssymb,graphicx,braket,bbold,mathtools,array}
\usepackage{multirow}
\usepackage[]{units}
\usepackage{hyperref}
\hypersetup{
    colorlinks=true,
    linkcolor=blue,
    filecolor=blue,      
    urlcolor=blue,
    citecolor=blue
}
\usepackage[utf8]{inputenc}
\usepackage[T1]{fontenc}

\newcommand{\cre}[1]{c_{#1}^\dagger}
\newcommand{\ann}[1]{c_{#1}^{}}

\newcommand{\ku}{K}
\newcommand{\kd}{\overline{K}}
\newcommand{\su}{\uparrow}
\newcommand{\sd}{\downarrow}
\newcommand{\fu}{T_{+}}
\newcommand{\fd}{T_{-}}
\newcommand{\fufd}{T_{\pm}}
\newcommand{\au}{S}
\newcommand{\ad}{T_0}
\newcommand{\nup}{n}
\newcommand{\nd}{\overline{n}}

\newcommand{\id}{\mathbb{1}}

\newcommand{\vect}[1]{\boldsymbol{#1}}

\newcommand{\where}[2]{\left . {#1} \right |_{#2}}

\newcommand{\h}[2]{h_{#1}^{#2}}
\newcommand{\dhS}[1]{\delta h_S^{#1}}
\newcommand{\dhV}[1]{\delta h_V^{#1}}
\newcommand{\dhl}[1]{\delta h_\ell^{#1}}
\newcommand{\HS}[1]{H_S^{#1}}
\newcommand{\HV}[1]{H_V^{#1}}
\newcommand{\Hl}[1]{H_\ell^{#1}}

\newcommand{\hamtot}{H_{\text{tot}}}
\newcommand{\hameff}{H_{\text{eff}}}
\newcommand{\hameffkramers}{H_{\text{eff, }\mathcal{N} \times \mathcal{N}}}
\newcommand{\hamhubb}{H_U}
\newcommand{\hamdetun}{H_\varepsilon}
\newcommand{\hamtunn}{H_t}
\newcommand{\hamspinsplit}{H_\Delta}
\newcommand{\hamspinsplitLR}{H_{\Delta_L,\Delta_R}}
\newcommand{\hamspinsplitgeneral}{H_{\Delta_L,\Delta_R,xyz}}
\newcommand{\hamspinmix}{H_S}
\newcommand{\hamvalleyzeeman}{H_V}

\begin{document}

  \title{Effective theory of monolayer TMDC double quantum dots}
  \author{Alessandro David}
  \email{alessandro.david@uni-konstanz.de}
  \author{Guido Burkard}
  \author{Andor Korm\'anyos}
  \affiliation{Department of Physics, University of Konstanz, D-78464, Germany}
  \date{\today}
  
  \begin{abstract}
    Monolayer Transition Metal Dichalcogenides (TMDCs) are promising candidates for quantum technologies, such as quantum dots, because they are truly two-dimensional semiconductors with a direct band gap. In this work, we analyse theoretically the behaviour of a double quantum dot (DQD) system created in the conduction band of these materials, with two electrons in the (1,1) charge configuration. Motivated by recent experimental progress, we consider several  scenarios, including different spin-orbit splittings in the two dots and including the case when the valley degeneracy is lifted due to an insulating ferromagnetic substrate. Finally, we discuss in which cases it is possible to reduce the low energy subspace to the lowest Kramers pairs. We find that in this case the low energy model is formally identical to the Heisenberg exchange Hamiltonian.
  \end{abstract}
  
  \maketitle
  
  \section{Introduction}
  
  Monolayers of transition metal dichalcogenides (TMDCs) are a class of 2D materials with very interesting electronic and optical properties \cite{wang_electronics_2012, choi_recent_2017}. They are atomically thin semiconductors with a direct band gap and two-fold degenerate valleys in the Brillouin zone \cite{mak_atomically_2010, splendiani_emerging_2010}. Early studies discovered that their intrinsic spin-orbit interaction splits the spin states in the valence band and that it is possible to optically manipulate the spin and valley degrees of freedom (DOF) \cite{xiao_coupled_2012, mak_control_2012}. Further theoretical works suggested the idea that the conduction band should be spin-split as well \cite{kormanyos_spin-orbit_2014, kosmider_large_2013} which found recent experimental confirmation \cite{wang_probing_2017}.
  
  The possibility to construct purely two-dimensional, electrostatically defined quantum dots (QDs) is one of the reasons that makes monolayer TMDCs so attractive, at least from a fundamental point of view. Compared to III-V semiconductors, such as GaAs \cite{petta_coherent_2005, hanson_spins_2007}, TMDCs have several isotopes with vanishing nuclear spin, thus lacking hyperfine interactions with the electronic spin. Moreover, TMDCs comprise an additional valley pseudospin. Although these two features are common to several other systems used for QDs, such as Si/SiGe quantum wells \cite{eriksson_spin-based_2004, zwanenburg_silicon_2013, knapp_characterization_2016, zajac_resonantly_2018}, graphene \cite{silvestrov_quantum_2007, trauzettel_spin_2007, guttinger_spin_2010, freitag_electrostatically_2016, freitag_tunable_2017} and carbon nanotubes (CNTs) \cite{buitelaar_multiwall_2002, graber_molecular_2006, kuemmeth_coupling_2008, abulizi_full_2016}, TMDCs are special because they exhibit very strong spin-orbit coupling (SOC).
  %
  % Silicon and carbon have nuclear spin-free stable isotopes (${}^{28}$Si, 92.23\%, which can be purified and ${}^{12}$C, 98.9\%), but their typical structures for QDs have an additional valley DOF, such as in Si/SiGe quantum wells, carbon nanotubes (CNT) and graphene \cite{novoselov_electric_2004, novoselov_two-dimensional_2005, castro_neto_electronic_2009}. Graphene being a semimetal the first attempts to confine electrons required the use of graphene nanoribbons \cite{silvestrov_quantum_2007, trauzettel_spin_2007, guttinger_spin_2010} which introduce scattering through disordered edges and only recently electrostatic confinement has been demonstrated with the electric potential of an STM tip over monolayer graphene and boron nitride substrate \cite{freitag_electrostatically_2016, freitag_tunable_2017}.
  %
  Theoretical investigation of QDs in TMDCs started with the magnetic field dependence of the single-electron spectrum \cite{kormanyos_spin-orbit_2014} and it now includes studies of valley hybridisation \cite{liu_intervalley_2014}, flake QDs of triangular and hexagonal shape \cite{pavlovic_electronic_2015}, the valley Zeeman effect \cite{dias_robust_2016}, optical control of a spin-valley qubit in nanostructures \cite{wu_spin-valley_2016}, spin-degenerate regimes for small QD radius in an external magnetic field \cite{brooks_spin-degenerate_2017}, a model of spin relaxation \cite{pearce_electron_2017}, electric control of a spin-valley qubit \cite{szechenyi_impurity-assisted_2017} and a model of valley qubit \cite{pawlowski_valley_2018}. On the experimental side, gating monolayer TMDCs is not straightforward \cite{lee_coulomb_2016, wang_electrical_2018} and low material quality has hindered for a long time the experimental study of the intrinsic properties of these materials. However, recently there has been a significant progress in the fabrication process of nanostructures in TMDCs. This has enabled the creation of single QDs on monolayer \cite{song_gate_2015, pisoni_gate-tunable_2018} or trilayer TMDCs \cite{wang_electrical_2018}, double QD experiments with tunable coupling strength between the dots \cite{zhang_electrotunable_2017, pisoni_gate-tunable_2018} and the observation of gate-controlled Coulomb blockade effect \cite{song_gate_2015, lee_coulomb_2016, zhang_electrotunable_2017, wang_electrical_2018, pisoni_gate-tunable_2018}.
  
  In this paper we study double quantum dots (DQDs) taking into account the spin-orbit coupling and both spin and valley DOF. We note that, although single \cite{kuemmeth_coupling_2008} and double \cite{von_stecher_double_2010, weiss_spin-orbit_2010, palyi_spin-valley_2010} quantum dots in CNTs have been studied, where the low energy theory shares some similarities with TMDCs, little is known about TMDC DQDs and the role of the exchange interaction. For simplicity we restrict each dot to the lowest orbital and we consider the system filled with two electrons. Starting from the model introduced in Ref.~\cite{rohling_universal_2012}, we add the spin-orbit interaction and we find a low energy effective Hamiltonian for the case where each dot is occupied by one electron. We also investigate the situations where the spin-orbit splitting is different for each dot and where we include the effects of a magnetic field. Furthermore, for large spin-orbit splittings it is convenient to focus on a smaller subset of states formed by the two lowest Kramers pairs of the system. In this smaller subspace we find that in most cases the interaction  is formally identical to the Heisenberg exchange interaction used to perform a CNOT gate between spin qubits \cite{loss_quantum_1998}. This supports the idea that the lowest Kramers pair can serve as a qubit for TMDC, as was suggested in Ref.~\cite{kormanyos_spin-orbit_2014}.
  
  This paper is organised as follows. In Sec.~\ref{sec:model} we introduce the basic terms appearing in our Hamiltonian, we present a naming conventions for certain useful projection operators and we briefly evoke the results for the null spin-orbit coupling case. Then, in Sec.~\ref{sec:oneonesub} we present the effective Hamiltonians for the case where each QD is occupied by one electron, for three different situations: when the spin-splitting is equal in both dots, when the spin-orbit splitting is different and when the TMDC is deposited on an insulating ferromagnetic substrate or placed in an external magnetic field. Afterwards, we explore in Sec.~\ref{sec:negativekramerspairs} under which conditions it is possible to focus on smaller effective Hamiltonians for these three situations. Finally, in Sec.~\ref{sec:conclusions} we present our conclusions.
  
  \section{Model}
  \label{sec:model}
  
  \subsection{Basic definitions}\label{sec:basicdefs}
  
  We give here the definitions for the different terms of the Hamiltonians that describe the various scenarios studied in this work. The operators $c_{j\tau\sigma}^{(\dagger)}$ annihilate (create) an electron in QD $j$ with valley $\tau$ and spin $\sigma$. Here $j=L$ ($R$) refers to the left (right) QD, $\tau = \ku$ ($\kd$) indicates the positive (negative) valley and $\sigma = \su$ ($\sd$) specifies spin up (spin down).
  
  The on-site Coulomb repulsion between electrons in the same QD is captured by the Hubbard Hamiltonian,
  \begin{equation}\label{eqn:hamhubb}
    \hamhubb = \frac{U}{2} \sum_{j = L,R} n_j (n_j - 1),
  \end{equation}
  where $U>0$ is the positive \emph{charging energy} of the dot and the number operator is defined as
  \begin{equation}
    n_j = \sum_{\tau = K, K'} \sum_{\sigma = \uparrow, \downarrow}
          \cre{j\tau\sigma} \ann{j\tau\sigma}.
  \end{equation}
  
  A detuning term specifies the energy difference $\varepsilon$ between the dots,
  \begin{equation}\label{eqn:hamdetun}
    \hamdetun = \frac{\varepsilon}{2} (n_L - n_R).
  \end{equation}
  
  Electron-hopping from one dot to the other is accounted for by a tunneling term that preserves spin and valley,
  \begin{equation}\label{eqn:hamtunn}
    \hamtunn = \sum_{\tau, \sigma}
    \left ( t \, \cre{R\sigma\tau} \ann{L\sigma\tau} + \text{h.c.} \right ),
  \end{equation}
  where the tunneling coefficient $t$ is generally a complex number.
  
  The intrinsic spin-orbit coupling is modeled by a simple time-reversal symmetric ($T$-symmetric) spin-splitting: $\Delta \tau_z \sigma_z$ \cite{kormanyos_spin-orbit_2014}. Here, $\tau_i$ ($\sigma_i$) is the $i$-th Pauli matrix acting on the valley (spin) DOF ($i = x, y, z$), while $\Delta$ is a real, positive or negative, coupling constant. This implies that the Kramers pair of states in the set $\mathcal{P} = \left \{ \ket{\ku\su}, \ket{\kd\sd} \right \}$ is shifted by the energy $+\Delta$, while the Kramers pair of states in the set $\mathcal{N} = \left \{ \ket{\ku\sd}, \ket{\kd\su} \right \}$ is shifted by the energy $-\Delta$. We call $\mathcal{P}$ the \emph{positive} Kramers pair and $\mathcal{N}$ the \emph{negative} Kramers pair. For the double dot system,
  \begin{equation}
    \label{eqn:spinsplitting}
    \begin{aligned}
      \hamspinsplit & = \Delta \sum_{j,\tau,\sigma} \cre{j\tau\sigma} (\tau_z)_{\tau\tau}  (\sigma_z)_{\sigma\sigma} \ann{j\tau\sigma}.
    \end{aligned}
  \end{equation}
  Eq.~\eqref{eqn:spinsplitting} assumes that the spin-orbit splitting is the same for every dot, which is usually the case for dots created on the same material. In case the spin-orbit splitting is different we use the following generalisation,
  \begin{equation}
    \label{eqn:spinsplittingLR}
    \begin{aligned}
      \hamspinsplitLR & = \sum_j \Delta_j \sum_{\tau,\sigma} \cre{j\tau\sigma} (\tau_z)_{\tau\tau} (\sigma_z)_{\sigma\sigma} \ann{j\tau\sigma},
      \end{aligned}
  \end{equation}
  where $\Delta_L$ and $\Delta_R$ are the spin-orbit splittings in the left and right QD respectively.
  
  We also consider the coupling of spin and valley to an external magnetic field. The corresponding spin Zeeman term is given by
  \begin{equation}\label{eqn:spinmixing}
    \begin{aligned}
      \hamspinmix & = \sum_j \vect{h}_{Sj} \cdot \sum_{\tau,\sigma_1,\sigma_2} \cre{j\tau\sigma_1} (\vect{\sigma})_{\sigma_1\sigma_2} \ann{j\tau\sigma_2},
    \end{aligned}
  \end{equation}
  where $\vect{h}_{SL}$ and $\vect{h}_{SR}$ are two vectors of coupling constants for left and right QD respectively and $\vect{\sigma}$ is the vector of Pauli matrices acting on spin.
  
\begin{figure}
    \centering
    \includegraphics{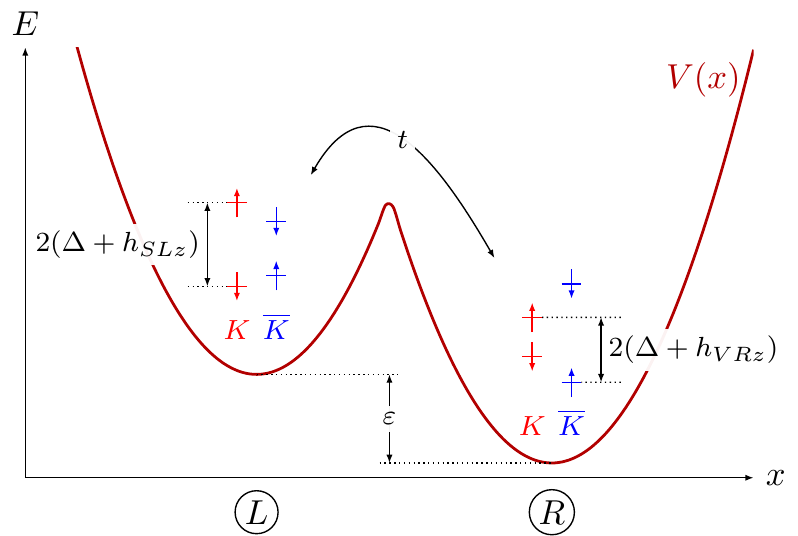}
    \caption{Diagram of the energy $E$ as a function of the position $x$ along the axis of the DQD. $V(x)$ (dark red line) represents the double-well potential that defines the left ($L$) and right ($R$) QDs. The energy levels of the valley and spin states inside the dots are shown here with a positive detuning $\varepsilon$. Spin states with valley $\ku$ ($\kd$) are coloured in red (blue). The energy levels are shifted by a symmetric spin-orbit splitting ($\Delta$) and by inhomogeneous spin and valley Zeeman terms along the $z$-direction, with coupling constants $h_{SLz}$/$h_{SRz}$ and $h_{VLz}$/$h_{VRz}$ respectively. Electrons are allowed to tunnel from one dot to the other with tunneling coefficient $t$.}
    \label{fig:energydiagram}
  \end{figure}

  The valley Zeeman term is the valley counterpart of the spin Zeeman but considering only the $z$-Pauli matrix acting on the valley. This is motivated by Ref.~\cite{kormanyos_spin-orbit_2014} and recent experiments \cite{zhao_enhanced_2017, zhong_van_2017},
  \begin{equation}\label{eqn:valleyzeeman}
      \hamvalleyzeeman = \sum_j h_{Vjz} \sum_{\tau,\sigma} \cre{j\tau\sigma} (\tau_z)_{\tau\tau} \ann{j\tau\sigma}
  \end{equation}
  where $h_{VLz}$ and $h_{VRz}$ describe the valley splittings in the left and right dot. A concise outline of the interactions and the DOF of our DQD model is shown in Fig.~\ref{fig:energydiagram}.
  
  We consider the case where there are two electrons in the system. The possible \emph{charge configurations} are $(2,0)$, $(1,1)$ and $(0,2)$, where $(n_L, n_R)$ means that there are $n_L$ electrons in the left QD and $n_R$ electrons in the right QD. Because of the spin and valley degrees of freedom and due to the Pauli exclusion principle, there is a total number of 28 linearly independent states: 6 $(2,0)$-states, 16 $(1,1)$-states and 6 $(0,2)$-states. Throughout this paper, we always assume a small detuning and weak tunneling, i.e.
  \begin{equation}\label{eqn:mainswcond}
    |t| \ll |U \pm \varepsilon|.
  \end{equation}
  
  \subsection{Naming basis states and projection operators}
  
  It is natural to refer to certain operators that will appear as projectors on specific states. We present a convention to name all 28 states of the total Hilbert space and we give the projection operators that will be useful later.
  
  \begin{table}
    \centering
    \begin{tabular}{c@{\hspace{.3cm}}|*{4}{@{\hspace{.3cm}}c}}
      & $\ket{S}$ & $\ket{T_-}$ & $\ket{T_0}$ & $\ket{T_+}$ \\[.1cm] \hline \\[-.2cm]
      Spin & $\frac{\ket{\su\sd}-\ket{\sd\su}}{\sqrt{2}}$
           & $\ket{\sd\sd}$
           & $\frac{\ket{\su\sd}+\ket{\sd\su}}{\sqrt{2}}$
           & $\ket{\su\su}$ \\[.2cm] \hline \\[-.2cm]
      Valley & $\frac{\ket{\ku\kd}-\ket{\kd\ku}}{\sqrt{2}}$
           & $\ket{\kd\kd}$
           & $\frac{\ket{\ku\kd}+\ket{\kd\ku}}{\sqrt{2}}$
           & $\ket{\ku\ku}$
    \end{tabular}
    \caption{Definitions of singlet and triplet states for the spin and valley DOF.}
    \label{tab:singlettripletstates}
  \end{table}
  
  Not all the states in the $(1,1)$-subspace are allowed to tunnel to $(0,2)$ or $(2,0)$-states, but only those which are antisymmetric in spin and valley, because of the Pauli exclusion principle \cite{palyi_hyperfine-induced_2009, palyi_spin-valley_2010}. Following the antisymmetric nature of the tunneling states, we begin by introducing a basis consisting of states that are symmetric or antisymmetric in both spin and valley:~$\ket{s_V s_S}_{(n_L, n_R)}$, where $(n_L, n_R)$ is the charge configuration and $s_V$ ($s_S$) indicates the exchange symmetry of valley (spin) DOF, which can be either a singlet ($S$) or a triplet ($T_-$, $T_0$, $T_+$). See Table~\ref{tab:singlettripletstates} for the definitions of singlet and triplet states for spin and valley. When $(n_L, n_R) = (1,1)$ we omit the indication of the charge configuration in the subscript. The equal spin-orbit coupling term $\hamspinsplit$ (Eq.~\eqref{eqn:spinsplitting}) is not diagonal in this basis. In order to work with a basis that makes $\hamspinsplit$ diagonal, in the $(1,1)$-subspace we substitute $\ket{S T_0}$, $\ket{T_0 S}$, $\ket{S S}$ and $\ket{T_0 T_0}$ with the following states,
  % \begin{subequations}
  %   \begin{align}
  %     \ket{\nup\nup} & = (\ket{S T_0} + \ket{T_0 S})/\sqrt{2}, \label{eqn:nupnup}\\
  %     \ket{\nd\nd} & = (\ket{S T_0} - \ket{T_0 S})/\sqrt{2}, \label{eqn:ndnd} \\
  %     \ket{\nup\nup}^+ & = (\ket{T_0 T_0} + \ket{S S})/\sqrt{2}, \\
  %     \ket{\nd\nd}^+ & = (\ket{T_0 T_0} - \ket{S S})/\sqrt{2},
  %   \end{align}
  % \end{subequations}
  \begin{subequations}
    \begin{align}
      \ket{\nup_\pm} & = (\ket{S T_0} \pm \ket{T_0 S})/\sqrt{2}, \\
      \ket{\nd_\pm} & = (\ket{T_0 T_0} \pm \ket{S S})/\sqrt{2},
    \end{align}
  \end{subequations}
  where $\ket{\nup_\pm}$ ($\ket{\nd_\pm}$) are antisymmetric (symmetric) spin-valley states consisting of only positive (subscript $+$) or negative (subscript $-$) Kramers pairs. Note that $\ket{\nup_\pm}$ ($\ket{\nd_\pm}$) are odd (even) under the time-reversal operator $T$. For these states there is no defined exchange symmetry for spin or valley alone. States analogous to $\ket{\nup_\pm}$ are defined in the $(2,0)$ and $(0,2)$-subspaces. In Table~\ref{tab:spinsplitshift}, we list all these states grouped by charge configuration and by symmetry under exchange.
  
  Projection operators on each one of the antisymmetric $(1,1)$-states are easy to obtain (see Appendix \ref{sec:projectors}),
  \begin{subequations}
    \label{eqn:antisymmprojectors}
    \begin{align}
      \label{eqn:projfufdau}
      P_{\ket{\fufd\au}} & = 
      \begin{multlined}[t][.7\columnwidth]
        \frac{1}{16}
            \left ( \id \pm \tau_{Lz} \right )
            \left ( \id \pm \tau_{Rz} \right )
            \left ( \id - \vect{\sigma}_{L} \cdot \vect{\sigma}_{R} \right ),
      \end{multlined} \\
      \label{eqn:projaufufd}
      P_{\ket{\au\fufd}} & = 
      \begin{multlined}[t][.7\columnwidth]
        \frac{1}{16}
            \left ( \id \pm \sigma_{Lz} \right )
            \left ( \id \pm \sigma_{Rz} \right )
            \left ( \id - \vect{\tau}_{L} \cdot \vect{\tau}_{R} \right ),
      \end{multlined} \\
      \label{eqn:projnup}
      P_{\ket{\nup_\pm}} & = 
      \begin{multlined}[t][.7\columnwidth]
        \frac{1}{16}
            \left ( \id - \tau_{Lz} \tau_{Rz} \right )
            \left ( \id - \sigma_{Lz} \sigma_{Rz} \right ) \\ \times
            \left ( \id \pm \tau_{Lz} \sigma_{Lz} \right )
            \left ( \id - \tau_{Lx} \sigma_{Lx} \tau_{Rx} \sigma_{Rx} \right ),
      \end{multlined}
    \end{align}
  \end{subequations}
  where $\tau_{ji}$ ($\sigma_{ji}$) is the $i$-th Pauli matrix acting on valley (spin) in QD $j = L,R$ and $\vect{\sigma}_j$, $\vect{\tau}_j$ are vectors of Pauli matrices acting on spin or valley respectively. We note that the projection operator over the whole antisymmetric subspace of the $(1,1)$-sector can be written as \cite{rohling_universal_2012} (see Appendix \ref{sec:projectors}),
  \begin{equation}\label{eqn:antisymmproj}
    P_{\text{as}} = (3 - \vect{\sigma}_L \cdot \vect{\sigma}_R - 
    \vect{\tau}_L \cdot \vect{\tau}_R - 
    (\vect{\sigma}_L \cdot \vect{\sigma}_R)(\vect{\tau}_L \cdot \vect{\tau}_R)) / 8.
  \end{equation}
  
  \subsection{Zero spin-orbit splitting ($\Delta = 0$)}\label{sec:zerospinsplitting}
  
  We finish this section by briefly considering the case where there is no spin-orbit splitting ($\Delta_L = \Delta_R = 0$). The total Hamiltonian is, then,
  \begin{equation}
    \hamtot = \hamhubb + \hamdetun + \hamtunn,
  \end{equation}
  where $\hamhubb$, $\hamdetun$ and $\hamtunn$ are defined in equations \eqref{eqn:hamhubb}, \eqref{eqn:hamdetun} and \eqref{eqn:hamtunn} respectively. This model describes DQD systems with fourfold degenerate spin and valley states in each dot (and no spin-orbit interaction). This case has been extensively treated in Ref.~\cite{rohling_universal_2012}. For later reference and readability we report the resulting effective Hamiltonian,
  \begin{equation}\label{eqn:heffexchangeonly}
    H_{\text{eff}} = -J P_{\text{as}},
  \end{equation}
  the \emph{exchange energy} $J$ in this case has the usual form,
  \begin{equation}\label{eqn:exchangeenergy}
    J = \frac{4 |t|^2 U}{U^2 - \varepsilon^2},
  \end{equation}
  and the definition of $P_{\text{as}}$ is given in Eq.~\eqref{eqn:antisymmproj}. This Hamiltonian shifts down the energy of all the (1,1) states in the antisymmetric subspace by $-J$. It means that the ground state is any state in this 6-dimensional subspace and it has energy $-J$. The first excited states are those in the orthogonal 10-dimensional symmetric subspace, with energy 0. This is represented in Fig.~\ref{fig:symmlevels}(a).
  
  \section{Results for (1,1)-subspace}\label{sec:oneonesub}
  
  \subsection{Symmetric spin-orbit splitting ($\Delta_L = \Delta_R = \Delta$)}
  
  We first consider the case where the spin-orbit coupling is equal for both QDs: $\Delta_L = \Delta_R = \Delta$. We call this \emph{symmetric} spin-orbit splitting. The total Hamiltonian is
  \begin{equation}\label{eqn:symmspinsplit}
    \hamtot = \hamhubb + \hamdetun + \hamtunn + \hamspinsplit,
  \end{equation}
  where $\hamspinsplit$ is defined in Eq.~\eqref{eqn:spinsplitting}. This Hamiltonian describes DQDs created in TMDCs \cite{mak_atomically_2010, splendiani_emerging_2010, mak_control_2012, xiao_coupled_2012, kormanyos_spin-orbit_2014, wang_probing_2017}, but it may also be used to study DQDs in CNTs \cite{ando_spin-orbit_2000, huertas-hernando_spin-orbit_2006, kuemmeth_coupling_2008, von_stecher_double_2010, weiss_spin-orbit_2010, palyi_spin-valley_2010}.
  
  Symmetric spin-orbit coupling $\hamspinsplit$ shifts those states formed by two elements of the negative (positive) Kramers pair by $-2\Delta$ ($+2\Delta$), while it leaves unchanged those states formed by one element of the negative Kramers pair and one element of the positive Kramers pair (see Table~\ref{tab:spinsplitshift}). In order to identify the $(1,1)$-sector as our \emph{low energy subspace} (LES) we have to guarantee first of all that no $(2,0)$ or $(0,2)$-state is lower in energy than any $(1,1)$-state. Looking at Table~\ref{tab:spinsplitshift} we see that this condition is met when
  \begin{equation}\label{eqn:symmspinsplitcond}
    4 |\Delta| < U-|\varepsilon|.
  \end{equation}
  When both \eqref{eqn:mainswcond} and \eqref{eqn:symmspinsplitcond} are satisfied, it follows that $(2,0)$ and $(0,2)$-states are energetically unfavored and tunneling out from the $(1,1)$-sector is strongly suppressed. However, virtual tunneling processes must be taken into account. It is important to notice that antisymmetric $(1,1)$-states can tunnel only to their $(0,2)$ and $(2,0)$ counterpart states that have the same spin and valley configuration. This is due to the spin- and valley-preserving nature of the tunneling term of Eq.~\eqref{eqn:hamtunn}: there is no transition from a negative Kramers pair to a positive Kramers pair and viceversa. Therefore, all the energy differences between initial and final states in a virtual tunneling process do not depend on $\Delta$ and the exchange interaction that emerges does not change from the case of zero spin-orbit splitting reported in Sec.~\ref{sec:zerospinsplitting}. Exchange interaction and symmetric spin-orbit coupling act independently of each other. We obtain the \emph{effective Hamiltonian}:
  \begin{equation}\label{eqn:heffsymmetric}
    \hameff = -JP_{\text{as}} + \Delta \Sigma,
  \end{equation}
  where $J$ is the same constant defined in \eqref{eqn:exchangeenergy}, $P_{\text{as}}$ is given in \eqref{eqn:antisymmproj}, $\Delta \Sigma = \where{\hamspinsplit}{(1,1)} = \Delta \left ( \tau_{Lz} \sigma_{Lz} + \tau_{Rz} \sigma_{Rz} \right )$ and $\where{\hamspinsplit}{(1,1)}$ is the restriction of $\hamspinsplit$ to the $(1,1)$-subspace.
  
  \begin{table}
    \centering
    \begin{tabular}{>{\centering}p{2cm}|>{\centering}p{2cm}>{\centering}p{2cm}>{\centering}p{2cm}}
      % \hline
      $\hamhubb + \hamdetun$ & $-2\Delta$ & $0$ & $+2\Delta$ \tabularnewline
      \hline
      \multicolumn{4}{c}{$(1,1)$-subspace, antisymmetric} \tabularnewline
      \hline
      \multirow{4}{*}{0} & $\ket{\nup_-}$ & $\ket{\fu\au}$ & $\ket{\nup_+}$ \tabularnewline
      & & $\ket{\fd\au}$ & \tabularnewline
      & & $\ket{\au\fu}$ & \tabularnewline
      & & $\ket{\au\fd}$ & \tabularnewline
      \hline
      \multicolumn{4}{c}{$(1,1)$-subspace, symmetric} \tabularnewline
      \hline
      \multirow{4}{*}{0} & $\ket{\nd_-}$ & $\ket{\fu\ad}$ & $\ket{\nd_+}$ \tabularnewline
      & $\ket{\fu\fd}$ & $\ket{\fd\ad}$ & $\ket{\fu\fu}$ \tabularnewline
      & $\ket{\fd\fu}$ & $\ket{\ad\fu}$ & $\ket{\fd\fd}$ \tabularnewline
      & & $\ket{\ad\fd}$ & \tabularnewline
      \hline
      \multicolumn{4}{c}{$(2,0)$-subspace} \tabularnewline
      \hline
      \multirow{4}{*}{$U+\varepsilon$} & $\ket{\nup_-}_{(2,0)}$ & $\ket{\fu\au}_{(2,0)}$ & $\ket{\nup_+}_{(2,0)}$ \tabularnewline
      & & $\ket{\fd\au}_{(2,0)}$ & \tabularnewline
      & & $\ket{\au\fu}_{(2,0)}$ & \tabularnewline
      & & $\ket{\au\fd}_{(2,0)}$ & \tabularnewline
      \hline
      \multicolumn{4}{c}{$(0,2)$-subspace} \tabularnewline
      \hline
      \multirow{4}{*}{$U-\varepsilon$} & $\ket{\nup_-}_{(0,2)}$ & $\ket{\fu\au}_{(0,2)}$ & $\ket{\nup_+}_{(0,2)}$ \tabularnewline
      & & $\ket{\fd\au}_{(0,2)}$ & \tabularnewline
      & & $\ket{\au\fu}_{(0,2)}$ & \tabularnewline
      & & $\ket{\au\fd}_{(0,2)}$ & \tabularnewline
      % \hline
    \end{tabular}
    \caption{The first column on the left hand side reports the value of Coulomb repulsion (Eq.~\eqref{eqn:hamhubb}) and detuning (Eq.~\eqref{eqn:hamdetun}) for the different charge configurations. The three columns on the right hand side show which states are shifted by $-2\Delta$, $0$ and $+2\Delta$ by the action of the symmetric spin-orbit coupling $\hamspinsplit$ defined in Eq.~\eqref{eqn:spinsplitting}. The states are grouped as $(1,1)$-states (antisymmetric and symmetric), $(2,0)$-states and $(0,2)$-states (only antisymmetric).}
    \label{tab:spinsplitshift}
  \end{table}
  
  \begin{figure}
    \centering
    \includegraphics{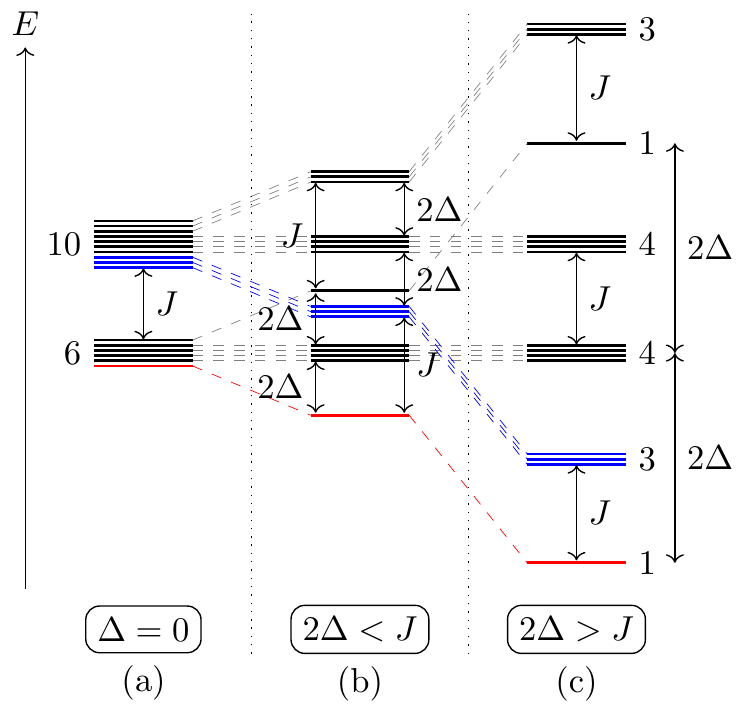}
    \caption{Level structure of $(1,1)$-states for the symmetric spin-orbit splitting case at fixed (small) detuning. We show three relevant cases with increasing spin-orbit splitting ($\Delta$) from left to right. Thick horizontal lines illustrate the energy levels. Degenerate levels are shown as a group of thick lines close together with a number indicating the degree of degeneracy. The coloured energy levels display states inside the $\mathcal{N} \times \mathcal{N}$-sector. The red energy levels indicate the antisymmetric state $\ket{\nup_-}$, while the blue energy levels are the symmetric states $\ket{\nd_-}$, $\ket{\fu\fd}$, $\ket{\fd\fu}$. In (a) the spin-orbit splitting is zero and there are only two degenerate energy levels, separated by the exchange energy $J$. The higher (lower) energy corresponding to the symmetric (asymmetric) states. In (b) both symmetric and antisymmetric energy levels are separated in three groups, $\ket{\nup_-}$ becomes the ground state and the first excited states are antisymmetric. In (c), when $2\Delta$ is greater than the exchange energy $J$, the $\mathcal{N} \times \mathcal{N}$-sector becomes the LES and the first excited states are symmetric.}
    \label{fig:symmlevels}
  \end{figure}
  
  It is simple to evaluate the consequences of Eq.~\eqref{eqn:heffsymmetric} since it is diagonal in the states of Table~\ref{tab:spinsplitshift}. Assuming now that $\Delta > 0$, the ground state is $\ket{\nup_-}$ because it is the only state that is shifted down by both $-J$ and $-2\Delta$, thus the ground state space is one dimensional, see Fig.~\ref{fig:symmlevels}(b),(c). The dimensionality of the ground state space changes from one to six only when $\Delta$ is exactly equal to zero (see Fig.~\ref{fig:symmlevels}(a)) and this degeneracy is lifted linearly in $\Delta$. As shown in Fig.~\ref{fig:symmlevels}(b),(c), there are five groups of excited states with various degrees of degeneracy and whose relative distances in energy depend on the sizes of $\Delta$ and $J$. For later comparison, we observe that in this case each antisymmetric state is separated from its symmetric counterpart by the same amount, the exchange energy $J$. In the case $\Delta < 0$, the results are analogous, but the ground state is $\ket{\nup_+}$.
  
  Regarding the experimental situation, the few available experiments indicate $U \approx 2$~meV for QDs created in WSe$_2$ \cite{song_gate_2015} and MoS$_2$ \cite{wang_electrical_2018, zhang_electrotunable_2017, pisoni_gate-tunable_2018} with QD radii around 100~nm. On the other hand, the theory predicts $2\Delta$ to be about 3~meV for MoS$_2$ and larger for other compounds \cite{kormanyos_spin-orbit_2014, kormanyos_k_2015}. Therefore, at the moment, condition \eqref{eqn:symmspinsplitcond} seems not to be satisfied for TMDCs, but further experiments with smaller QD sizes may lead to higher charging energies. However, when Eq.~\eqref{eqn:symmspinsplitcond} is not satisfied, it is possible to focus on a smaller subspace inside the $(1,1)$-sector, as explained in Sec.~\ref{sec:negativekramerspairs}.
  
  \subsection{Asymmetric spin-orbit splitting ($\Delta_L \neq \Delta_R$)}\label{sec:asymmspinsplit}
  
  Suppose now that the spin-orbit strength is different in the two dots, $\Delta_L \neq \Delta_R$, this is the case of \emph{asymmetric} spin-orbit splitting. The total Hamiltonian is
  \begin{equation}\label{eqn:asymmspinsplit}
    \hamtot = \hamhubb + \hamdetun + \hamtunn + \hamspinsplitLR,
  \end{equation}
  where $\hamspinsplit$ has been substituted by $\hamspinsplitLR$ defined in Eq.~\eqref{eqn:spinsplittingLR}.
  
  \begin{figure}
    \centering
    \includegraphics{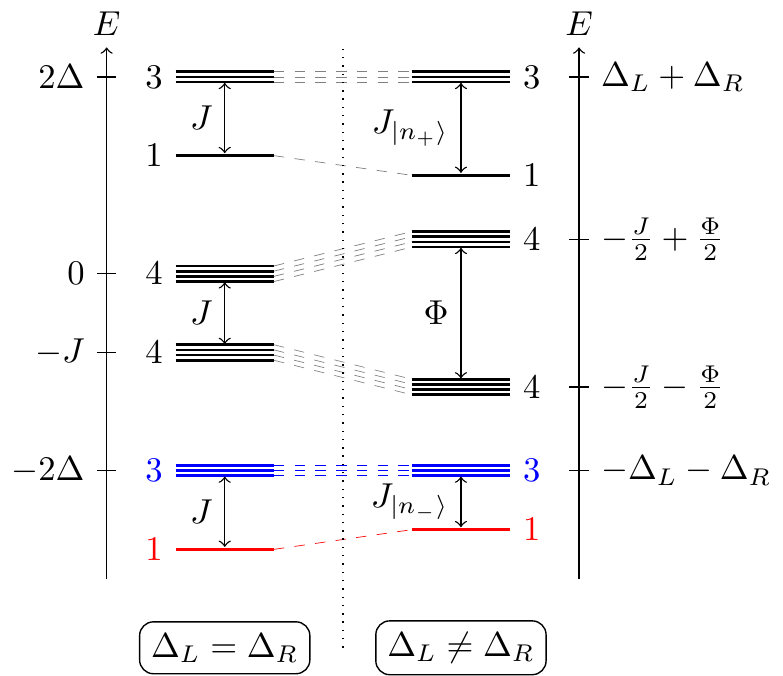}
    \caption{Comparison of the energy levels for symmetric (left) and asymmetric (right) spin-orbit splitting cases. The energy levels from left to right are aligned in a way to preserve the sum $\Delta_L + \Delta_R$ between the two cases (equal to $2\Delta$ for the symmetric spin-orbit splitting case). The left hand side is equivalent to the case of $2\Delta > J$ depicted in Fig.~\ref{fig:symmlevels}(c). The exchange energies $J$ and $J_{\ket{\nup_\pm}}$ are defined in Eq.~\eqref{eqn:exchangeenergy} and Eq.~\eqref{eqn:asymmexchangeenergy}, while $\Phi = \sqrt{J^2 + 4(\Delta_L - \Delta_R)^2}$, see Eq.~\eqref{eqn:offdiageigenval}.}
    \label{fig:asymmlevels}
  \end{figure}
  
  There is theoretical evidence that for (rather small) QDs on TMDCs the spin-orbit strength $\Delta$ depends on the radius of the QD \cite{brooks_spin-degenerate_2017}. Therefore, the Hamiltonian in Eq.~\eqref{eqn:asymmspinsplit} can reflect a situation where the two dots have different sizes. Changing the size of the QD would allow for a smooth and tunable adjustment of $\Delta$. Another situation where $\Delta_L \neq \Delta_R$ can be relevant is when the dots would be created in lateral heterojunctions such that each dot is created in different types of TMDC, which intrinsically possess different spin-orbit splittings \cite{kormanyos_k_2015}. Lateral heterojunctions of different TMDCs have already been demonstrated by several groups \cite{huang_lateral_2014, duan_lateral_2014, gong_vertical_2014}. To our knowledge no experiment has been reported yet involving such a DQD. We notice that for heterojunctions the situation might be further complicated by mismatch in the energy band gap, acting as a detuning away from the interface, and by likely non-negligible differences in the Coulomb charging energy $U$ of the two compounds.
  
  We point out that $\hamspinsplitLR$ is not diagonal in the basis of the states listed in Table~\ref{tab:spinsplitshift}. Each of the antisymmetric states $\ket{\fu\au}$, $\ket{\fd\au}$, $\ket{\au\fu}$, $\ket{\au\fd}$ is mixed with its symmetric counterpart and the off-diagonal elements are $\pm (\Delta_L - \Delta_R)$. This is relevant to determine the ground state, but not for the form of the effective Hamiltonian, because the mixing happens inside the $(1,1)$-subspace. Another difference is that, in contrast to Table~\ref{tab:spinsplitshift}, the $\pm 2\Delta$ on the diagonal of the Hamiltonian are replaced by $\pm (\Delta_L + \Delta_R)$ in the $(1,1)$-subspace, by $\pm 2\Delta_L$ in $(2,0)$ and by $\pm 2\Delta_R$ in $(0,2)$. When condition \eqref{eqn:mainswcond} is also valid, we can ensure that $(1,1)$-states are lower in energy than $(2,0)$ and $(0,2)$ states by requiring that
  \begin{multline}
    \max\{|\Delta_L + \Delta_R|, |\Delta_L - \Delta_R|\} + \\
    + 2 \max\{|\Delta_L|, |\Delta_R|\} < U - |\varepsilon|.
  \end{multline}
  In this case we can use the Schrieffer-Wolff transformation \cite{schrieffer_relation_1966, winkler_spin--orbit_2003, bravyi_schriefferwolff_2011} to obtain the effective Hamiltonian. The term $\hamspinsplitLR$ has three significant consequences. Firstly, the following matrix element differences now depend on $\pm (\Delta_L - \Delta_R)$,
  \begin{subequations}\label{eqn:asymmenergydiff}
    \begin{align}
        E_{\ket{\nup_\pm}} - E_{\ket{\nup_\pm}_{(0,2)}} & = \pm (\Delta_L - \Delta_R) - U + \varepsilon, \\
        E_{\ket{\nup_\pm}} - E_{\ket{\nup_\pm}_{(2,0)}} & = \mp (\Delta_L - \Delta_R) - U - \varepsilon,
      \end{align}
  \end{subequations}
  where $E_{\ket{\psi}} = \bra{\psi}\hamtot\ket{\psi}$. Secondly, in order to apply perturbation theory to the $(1,1)$-subspace, the coupling constants must satisfy the conditions $|t| \ll | U \pm \varepsilon |$, $|t| \ll | U \pm (\varepsilon + \Delta_L - \Delta_R) |$ and $|t| \ll | U \pm (\varepsilon - \Delta_L + \Delta_R) |$ simultaneously. In turn, these imply that $| \Delta_L - \Delta_R | \ll U$. Lastly, the exchange energies relative to $\ket{\nup_\pm}$ acquire a dependence on the asymmetry of the spin-orbit splittings,
  \begin{equation}\label{eqn:asymmexchangeenergy}
    J_{\ket{\nup_\pm}} = \frac{4 |t|^2 U}{(U + \varepsilon \pm \Delta_L \mp \Delta_R)
      (U - \varepsilon \mp \Delta_L \pm \Delta_R)},
  \end{equation}
  c.f. Eq.~\eqref{eqn:exchangeenergy}. The exchange energies for states $\ket{\fu\au}$, $\ket{\fd\au}$, $\ket{\au\fu}$ and $\ket{\au\fd}$ are not affected.
  
  An effective Hamiltonian can be written in a compact form in the following way,
  \begin{multline}\label{eqn:heffasymm}
    \hameff = -J P_{\text{as}} + \Delta_L \Sigma_L + \Delta_R \Sigma_R + \\
      - (J_{\ket{\nup_-}} - J) P_{\ket{\nup_-}} 
      - (J_{\ket{\nup_+}} - J) P_{\ket{\nup_+}},
  \end{multline}
  where $P_{\ket{\nup_-}}$ and $P_{\ket{\nup_+}}$ are the projectors on $\ket{\nup_-}$ and $\ket{\nup_+}$ defined in equation \eqref{eqn:projnup} and in this case we write the restriction $\where{\hamspinsplitLR}{(1,1)} = \Delta_L \Sigma_L + \Delta_R \Sigma_R,$ with $\Sigma_j = \tau_{jz} \sigma_{jz}$. By diagonalizing the Hamiltonian~\eqref{eqn:heffasymm}, we obtain the energy levels shown in Fig.~\ref{fig:asymmlevels}. Interestingly, the two levels that appear at energy 0 and energy $-J$ for the symmetric spin-orbit splitting case move to the energies,
  \begin{equation}\label{eqn:offdiageigenval}
    -\frac{J}{2} \pm \frac{1}{2} \sqrt{J^2 + 4(\Delta_L - \Delta_R)^2}
    = -\frac{J}{2} \pm \frac{\Phi}{2}.
  \end{equation}
  This is a consequence of the level repulsion due to the off-diagonal $\pm(\Delta_L - \Delta_R)$ matrix elements coming from $\hamspinsplitLR$. However, the degrees of degeneracy remain the same. Looking at Eq.~\eqref{eqn:offdiageigenval} and at Fig.~\ref{fig:asymmlevels}, we realise that, for positive $\Delta_L+\Delta_R$, the ground state is $\ket{\nup_-}$, unless $|\Delta_L - \Delta_R|$ is larger than $\Delta_L + \Delta_R$, in which case the states $\ket{\fu\au}$, $\ket{\fd\au}$, $\ket{\au\fu}$, $\ket{\au\fd}$ (or their symmetric counterparts) span a 4-fold degenerate ground state space.
  
  We want to briefly mention here that a generalisation of the asymmetric spin-orbit coupling which preserves the $T$-symmetry, including couplings to the in-plane components of the spins, does not affect the exchange interaction. Consider
  \begin{equation}\label{eqn:generalspinsplitting}
      H_{\Delta_L, \Delta_R, xyz} = \sum_j \vect{\Delta}_j \cdot
      \sum_{\tau, \sigma_1, \sigma_2} \cre{j\tau\sigma_1} (\tau_z)_{\tau\tau}
      (\vect{\sigma})_{\sigma_1\sigma_2} \ann{j\tau\sigma_2}
  \end{equation}
  where $\vect{\Delta}_j$ is the vector of the three coupling constants $\Delta_{jx}$,  $\Delta_{jy}$,  $\Delta_{jz}$ for QD $j$, multiplied by the vector of spin Pauli matrices $\vect{\sigma}$. Eq.~\eqref{eqn:generalspinsplitting} resembles the spin Zeeman term (cf. Eq.~\eqref{eqn:spinmixing}), but it also preserves $T$-symmetry because of the $\tau_z$ that multiplies the spin Pauli matrices. Qualitatively, a term like this should appear in the Hamiltonian of TMDC DQD systems for which it is possible to define an average local tilt and an average local curvature of the TMDC sheet for each dot \cite{pearce_tight-binding_2016}. However, tilt and curvature add couplings between conduction and valence band as well and one would need to check whether these couplings are small enough to be neglected. This is not the focus of this work and we only want to stress that substituting $\hamspinsplitLR$ with $\hamspinsplitgeneral$ in Eq.~\eqref{eqn:asymmspinsplit} does not modify the exchange interaction given in Eq.~\eqref{eqn:heffasymm}.
  
  \subsection{Spin and valley Zeeman and symmetric spin-orbit coupling}\label{sec:spinvalleyzeemanoneone}
  
  In this section we investigate the influence of spin and valley Zeeman terms on the symmetric spin-orbit splitting case. The total Hamiltonian for this scenario reads
  \begin{equation}\label{eqn:spinmixingvalleyzeeman}
    \hamtot = \hamhubb + \hamdetun + \hamtunn + \hamspinsplit + \hamspinmix
    + \hamvalleyzeeman,
  \end{equation}
  where $\hamspinmix$ and $\hamvalleyzeeman$ are defined by equations \eqref{eqn:spinmixing} and \eqref{eqn:valleyzeeman} respectively. This is the case depicted in Fig.~\ref{fig:energydiagram}.
  
  This model approximates at least two similar but distinct situations. First, when a monolayer TMDC is placed in a non-uniform and weak magnetic field, the spin of the electron in each dot couples to the local field in all three spacial directions through the spin Zeeman interaction $\hamspinmix$. On the other hand, we assume that the valley only couples along $z$, the orthogonal direction to the plane \cite{kormanyos_spin-orbit_2014}. In contrast to Ref.~\cite{kormanyos_spin-orbit_2014} here we only consider the lowest orbital state in each of the dots but we also take into account the Coulomb charging energy $U$. The other possible situation that came to our attention is that of monolayer TMDC deposited on a ferromagnetic insulator, e.g. europium oxide (EuO). Such a situation was considered in Ref.~\cite{qi_giant_2015}, where Eq.~\eqref{eqn:spinmixing} was used to model the giant and tunable valley splitting. A Rashba term is also included, mixing conduction and valence band, that we do not consider here. Although Ref.~\cite{qi_giant_2015} does not consider the valley Zeeman term, we note that it is allowed in the effective Hamiltonian of the TMDC because the ferromagnetic substrate breaks time-reversal symmetry. The presence of such magnetic exchange field in TMDC and ferromagnetic semiconductor heterostructures has been recently demonstrated \cite{zhao_enhanced_2017, zhong_van_2017}.
  
  Now both $\hamspinmix$ and $\hamvalleyzeeman$ are non-diagonal in the basis of states presented in Table~\ref{tab:spinsplitshift} and they mix a large number of states. We do not list all the newly introduced matrix elements but we observe that they are confined inside the charge sectors $(0,2)$, $(1,1)$ and $(2,0)$, as $\hamspinmix$ and $\hamvalleyzeeman$ do not contain terms of the form $\cre{L\tau\sigma}\ann{R\tau'\sigma'}$ (or h.c.). The only coupling between $(1,1)$-states and $(2,0)$ or $(0,2)$-states remains tunneling. Nevertheless, we have to ensure that $(1,1)$-states are lower in energy than other states. In Appendix \ref{sec:lesconditions} we report the discussion of which conditions must be fulfilled in order for this to be the case. In addition, to apply the Schrieffer-Wolff transformation, Eq.~\eqref{eqn:mainswcond} and the conditions
  \begin{subequations}
    \begin{align}
      |t| & \ll U \pm (\varepsilon + (h_{\ell Lz} - h_{\ell Rz})), \\
      |t| & \ll U \pm (\varepsilon - (h_{\ell Lz} - h_{\ell Rz})),
    \end{align}
  \end{subequations}
  where $\ell=V,S$, must also be satisfied. The matrix element differences of states $\ket{\fd\au}$, $\ket{\fu\au}$, $\ket{\au\fd}$, $\ket{\au\fu}$ with their $(2,0)$ and $(0,2)$ counterparts acquire a dependence on $h_{VLz} - h_{VRz}$ or $h_{SLz} - h_{SRz}$. Defining the exchange energies
  \begin{subequations}
    \begin{align}
      J_{\ket{\fufd\au}} & = \frac{4 |t|^2 U}{(U + \varepsilon \pm h_{VLz} \mp h_{VRz})
        (U - \varepsilon \mp h_{VLz} \pm h_{VRz})}, \\
      J_{\ket{\au\fufd}} & = \frac{4 |t|^2 U}{(U + \varepsilon \pm h_{SLz} \mp h_{SRz})
        (U - \varepsilon \mp h_{SLz} \pm h_{SRz})},
    \end{align}
  \end{subequations}
  the effective Hamiltonian can be written as
  \begin{multline}
    \hameff = -J (P_{\ket{\nup_-}} + P_{\ket{\nup_+}})
    -J_{\ket{\fd\au}} P_{\ket{\fd\au}}
    -J_{\ket{\fu\au}} P_{\ket{\fu\au}} + \\
    -J_{\ket{\au\fd}} P_{\ket{\au\fd}}
    -J_{\ket{\au\fu}} P_{\ket{\au\fu}}
    + \Delta \Sigma + \\
    + \vect{h}_{SL} \cdot \vect{\sigma}_{L}
    + \vect{h}_{SR} \cdot \vect{\sigma}_{R}
    + h_{VLz} \tau_{Lz} + h_{VRz} \tau_{Rz},
  \end{multline}
  where $J$ is the standard exchange energy defined in Eq.~\eqref{eqn:exchangeenergy} and $\vect{h}_{SL} \cdot \vect{\sigma}_{L} + \vect{h}_{SR} \cdot \vect{\sigma}_{R} \equiv \where{\hamspinmix}{(1,1)}$ while $h_{VLz} \tau_{Lz} + h_{VRz} \tau_{Rz} \equiv \where{\hamvalleyzeeman}{(1,1)}$.
  
  \section{Results for negative Kramers pairs subspace}
  \label{sec:negativekramerspairs}
  
  In this section we explain in which cases it is possible to restrict the LES to the tensor product of the two lowest Kramers pairs in energy and we present the effective Hamiltonians in this smaller subspace. In general, this happens when the spin-orbit strength is quite large, thus the results of this section are relevant for TMDCs. From now on we assume $\Delta_{L,R} > 0$, so that for each dot the lowest Kramers pair is $\mathcal{N}$, the negative one (see Sec.~\ref{sec:basicdefs}). Similar results can be obtained in case the spin-orbit splittings are negative and the lowest Kramers pair is $\mathcal{P}$, the positive one.
  
  \subsection{Symmetric spin-orbit splitting ($\Delta_L = \Delta_R = \Delta$)}
  
  In the case of symmetric spin-orbit splitting ($\Delta_L = \Delta_R = \Delta$, see $\hamtot$ in Eq.~\eqref{eqn:symmspinsplit}) there is no matrix element that allows a transition between $\mathcal{N}$ and $\mathcal{P}$. Assuming Eq.~\eqref{eqn:mainswcond} and $2\Delta > J$ (which is usually the case for TMDCs at low detuning) we see from Fig.~\ref{fig:symmlevels}(c) that the subspace spanned by the states of $\mathcal{N} \times \mathcal{N}$ (also called $\mathcal{N} \times \mathcal{N}$-sector) is the LES, where,
  \begin{multline}
    \mathcal{N} \times \mathcal{N} = \left \{ 
      \ket{\ku\sd ; \ku\sd}, \quad
      \ket{\ku\sd ; \kd\su}, \right . \\ \left .
      \ket{\kd\su ; \ku\sd}, \quad
      \ket{\kd\su ; \kd\su}
    \right \},
  \end{multline}
  with $\ket{\tau_1\sigma_1 ; \tau_2\sigma_2} = \cre{L\tau_1\sigma_1} \cre{R\tau_2\sigma_2} \ket{0}$. Now we present the effective Hamiltonian when the system is restricted to this LES.
  
  We can easily identify the negative Kramers pair $\mathcal{N}$ as a spin-1/2 DOF with
  \begin{equation}\label{eqn:newspins}
    \ket{\widetilde{\su}} \equiv \ket{\kd\su}, \qquad
    \ket{\widetilde{\sd}} \equiv \ket{\ku\sd},
  \end{equation}
  so that the effective basis $\mathcal{N} \times \mathcal{N}$ can be seen as the basis of all the states of two spins,
  \begin{equation}
    \mathcal{N} \times \mathcal{N} = \left \{ 
      \ket{\widetilde{\su} ; \widetilde{\su}}, \quad
      \ket{\widetilde{\su} ; \widetilde{\sd}}, \quad
      \ket{\widetilde{\sd} ; \widetilde{\su}}, \quad
      \ket{\widetilde{\sd} ; \widetilde{\sd}}
    \right \}.
  \end{equation}
  In this basis, the spin-orbit coupling always assumes the value of $-2\Delta$, which we ignore in the following since it is just an energy shift. It is well known that the effective Hamiltonian for two electrons with only the spin DOF in two QDs with only tunneling, detuning and Hubbard potential is given by \cite{hanson_spins_2007} $H_{\text{eff}} = -J \vect{S}_L \cdot \vect{S}_R$, with the usual exchange energy $J$ multiplied by the projector on the singlet state $\ket{S}$, the only antisymmetric state for two spins ($\vect{S}_j$ is the vector of spin operators acting on QD $j$). Analogously, using the following operators,
  \begin{equation}
    \begin{gathered}
      \widetilde{\sigma}_z = \sigma_z, \qquad
      \widetilde{\sigma}_x = \tau_x\sigma_x, \qquad
      \widetilde{\sigma}_y = \tau_x\sigma_y,
    \end{gathered}
  \end{equation}
  the Hamiltonian of the system restricted to basis $\mathcal{N} \times \mathcal{N}$ is
  \begin{equation}\label{eqn:hameffhighsymmsplit}
    \hameffkramers = 
    -J \widetilde{\vect{S}}_L \cdot \widetilde{\vect{S}}_R,
  \end{equation}
  where $\widetilde{S}_{ji} = \frac{1}{2}\widetilde{\sigma}_{ji}$ is the spin operator proportional to the new Pauli operator $\widetilde{\sigma}_i$, $i = x, y, z$, acting on QD $j = L, R$. The ground state space is one-dimensional as for the full $(1,1)$-subspace. Indeed, the ground state is the same,
  \begin{multline}
    \ket{\widetilde{S}} = 
      \left ( \ket{\widetilde{\su} ; \widetilde{\sd}} -
      \ket{\widetilde{\sd} ; \widetilde{\su}} \right ) / \sqrt{2} = \\
      \left ( \ket{\ku \sd ; \kd \su} -
      \ket{\kd \su ; \ku \sd} \right ) / \sqrt{2} = \ket{\nup_-}.
  \end{multline}
  
  \subsection{Almost symmetric spin-orbit splitting ($\left | \Delta_L - \Delta_R \right | \ll \Delta_L + \Delta_R$)}
  
  Very similar considerations are valid when we relax the constraint of equal spin-orbit splittings in the dots ($\Delta_L \neq \Delta_R$, see $\hamtot$ in \eqref{eqn:asymmspinsplit}). To reduce the LES to the subspace spanned by $\mathcal{N} \times \mathcal{N}$ we need
  \begin{equation}
    \Delta_L + \Delta_R > (J+\sqrt{J^2 + 4(\Delta_L - \Delta_R)^2})/2,
  \end{equation}
  see Fig.~\ref{fig:asymmlevels} and Eq.~\eqref{eqn:offdiageigenval}. Of course, to apply perturbation theory, the conditions $|t| \ll | U \pm (\varepsilon + \Delta_L - \Delta_R) |$ and $|t| \ll | U \pm (\varepsilon - \Delta_L + \Delta_R) |$ must be valid here too.
  
  Following the same steps that led to Eq.~\eqref{eqn:hameffhighsymmsplit}, we arrive to the effective Hamiltonian,
  \begin{equation}\label{eqn:heffkramersasymm}
    \hameffkramers = 
    -J_{\ket{\nup_-}} \widetilde{\vect{S}}_L \cdot \widetilde{\vect{S}}_R.
  \end{equation}
  where $J_{\ket{\nup_-}}$ is given in Eq.~\eqref{eqn:asymmexchangeenergy}. Indeed, from Fig.~\ref{fig:asymmlevels} we see that the energy levels for the $\mathcal{N} \times \mathcal{N}$-sector are similar between symmetric and asymmetric spin-orbit splitting, only the exchange energy $J$ is altered.
  
  To better understand what is the effect of $\Delta_L - \Delta_R$ in $J_{\ket{\nup_-}}$ with respect to $J$, we may write
  \begin{equation}
    J_{\ket{\nup_-}} = \frac{J}{\left ( 1 + \frac{\Delta_R - \Delta_L}{U + \varepsilon} \right ) \left ( 1 + \frac{\Delta_L - \Delta_R}{U - \varepsilon} \right )}.
  \end{equation}
  Since $(\Delta_L - \Delta_R)/(U \pm \varepsilon) \ll 1$, we can write
  \begin{equation}
    J_{\ket{\nup_-}} \simeq J \left ( 1 - 2 \varepsilon \frac{\Delta_L - \Delta_R}{U^2 - \varepsilon^2} \right ).
  \end{equation}
  The effect of an almost symmetric spin-orbit coupling on the exchange energy is that of fine tuning around the value of $J$ by a small positive or negative quantity, depending on the sign of $\varepsilon$ and $\Delta_L - \Delta_R$. Moreover, as $\varepsilon \to 0$, $J_{\ket{\nup_-}} \to J$. The above results remain valid for the generalisation to $\hamspinsplitgeneral$ (Eq.~\eqref{eqn:generalspinsplitting}).
  
  \subsection{Spin and valley Zeeman and symmetric spin-orbit coupling}
  
  We now consider the total Hamiltonian of Eq.~\eqref{eqn:spinmixingvalleyzeeman} restricted to the $\mathcal{N} \times \mathcal{N}$-sector. The results in this section are valid when the spin and valley Zeeman coupling constants are weak compared to the spin-orbit strength: $|h_{Sji}|, |h_{Vjz}| \ll 2\Delta$, $j = L, R$, $i = x, y, z$. We show directly and discuss the effective Hamiltonian for the subspace spanned by $\mathcal{N} \times \mathcal{N}$ as a $4 \times 4$ matrix. The columns are associated, from left to right, to the rotated basis states $\ket{\nup_-}$, $\ket{\nd_-}$, $\ket{\fu\fd}$ and $\ket{\fd\fu}$,
  \begin{widetext}
    \begin{equation}\label{eqn:heffkramersspinmixingvalleyzeeman}
        \hameffkramers =
        \begin{pmatrix}
          -J - A & \dhS{z} - \dhV{z} & 0 & 0 \\
          \dhS{z} - \dhV{z} & -A & 0 & 0 \\
          0 & 0 & - \HS{z} + \HV{z} - A_{+} & 0 \\
          0 & 0 & 0 & \HS{z} - \HV{z} - A_{-}
        \end{pmatrix},
      \end{equation}
  \end{widetext}
  where $J$ is the standard exchange energy as in Eq.~\eqref{eqn:exchangeenergy}. We use the notation
  \begin{subequations}\label{eqn:spinvalleynotation}
    \begin{align}
      \Hl{z} & = \h{\ell Lz}{} + \h{\ell Rz}{}, \\
      % \HV{z} & = \h{VLz}{} + \h{VRz}{}, \\
      \dhl{z} & = \h{\ell Lz}{} - \h{\ell Rz}{}, \\
      % \dhV{z} & = \h{VLz}{} - \h{VRz}{}, \\
      \h{Sj}{\pm} & = \h{Sjx}{} \pm i \h{Sjy}{},
    \end{align}
  \end{subequations}
  with $\ell = S,V$, $j=L,R$ and $A_{\pm}$ and $A$ are given by
  \begin{subequations}
    \begin{align}
      A_{\pm} & = \frac{\h{SL}{+}\h{SL}{-} + \h{SR}{+}\h{SR}{-}}{2\Delta \pm \HS{z}}, \\
      A & = \frac{A_{-} + A_{+}}{2} = 2\Delta \frac{\h{SL}{+}\h{SL}{-} + \h{SR}{+}\h{SR}{-}}{4\Delta^2 - {(\HS{z})}^2}.
    \end{align}
  \end{subequations}
  Since $|h_{Sjz}| \ll 2\Delta$, we can use the approximation $A_{+} \approx A_{-} \approx A$, thus the contribution of $A$ to Eq.~\eqref{eqn:heffkramersspinmixingvalleyzeeman} is an energy shift that can be ignored. Therefore, in this case we can also write
  \begin{equation}
    \hameffkramers = -J \widetilde{\vect{S}}_L \cdot \widetilde{\vect{S}}_R + \where{\hamspinmix}{\mathcal{N} \times \mathcal{N}} + \where{\hamvalleyzeeman}{\mathcal{N} \times \mathcal{N}},
  \end{equation}
  where $\where{\hamspinmix}{\mathcal{N} \times \mathcal{N}}$ and $\where{\hamvalleyzeeman}{\mathcal{N} \times \mathcal{N}}$ are the restrictions of $\hamspinmix$ and $\hamvalleyzeeman$ to the $\mathcal{N} \times \mathcal{N}$-sector.
  
  Given the generality of $\hamspinsplitgeneral$ (Eq.~\eqref{eqn:generalspinsplitting}) as a $T$-symmetric term and the fact that $\hamspinsplitgeneral$ and $\hamspinmix$ only differ in the presence of $\tau_z$ that multiplies the spin Pauli matrices, we can assert that the new matrix elements ($\HS{z}$, $\HV{z}$, $\dhS{z}$, $\dhV{z}$, $A_{\pm}$ and $A$) that appear in Eq.~\eqref{eqn:heffkramersspinmixingvalleyzeeman} as compared to Eq.~\eqref{eqn:hameffhighsymmsplit}, are allowed only by breaking the time-reversal symmetry.
  
  \section{Conclusions}\label{sec:conclusions}
  
  In this work we have studied and presented the influence of a spin-orbit coupling on the low energy properties of a DQD system with spin and valley DOF in the $(1,1)$ charge configuration. In our analysis we have also explored the possibility of a different spin-orbit splitting in each dot and we have included a $T$-symmetry breaking magnetic field. In addition, we have discussed under which conditions the LES corresponding to the $(1,1)$ charge configuration, which is 16-dimensional, can be further restricted to a 4-dimensional subspace (the $\mathcal{N} \times \mathcal{N}$-sector). We found that an equal spin-orbit splitting in each dot has no effects on the induced exchange interaction with respect to the case without spin-orbit coupling. On the other hand, asymmetric spin-orbit splitting, spin Zeeman and valley Zeeman modify the exchange coupling constants of three different pairs of antisymmetric states respectively. The modification of the exchange energies is similar for all these three pairs of states and it depends on the asymmetry of the interaction between left and right QD.
  
  We also found that TMDCs satisfy the conditions to restrict the LES to the $\mathcal{N} \times \mathcal{N}$-sector, where the effective Hamiltonian for the symmetric spin-orbit splitting case is formally identical to the Heisenberg exchange interaction between two spin-only qubits in valley non-degenerate materials. This renders the Kramers pair an ideal implementation of a qubit in TMDC, as was suggested in Ref.~\cite{kormanyos_spin-orbit_2014}. If the $\tau_x$ operation can be effectively implemented (theoretical proposals to achieve this include the use of impurities \cite{palyi_disorder-mediated_2011, szechenyi_impurity-assisted_2017} or the use of oscillating confinement potentials \cite{pawlowski_valley_2018}), a recipe for a CNOT gate with these states is readily available from the original Loss and DiVincenzo proposal for spin-only qubits \cite{loss_quantum_1998}. Moving to the asymmetric spin-orbit splitting we found that only the exchange energy is affected, while the form of the exchange Hamiltonian remains unchanged. The spin-orbit coupling asymmetry offers a way to tune the exchange energy other than the detuning of the dots. Finally, for the spin and valley Zeeman case, new couplings appeared in the reduced effective Hamiltonian, which originate from the breaking of the time-reversal symmetry.
  
  \section*{Acknowledgements}
  
  We thank M. Brooks and V. Shkolnikov for helpful discussions and proofreading. We acknowledge funding from FLAG-ERA through project “iSpinText” and from the Konstanz Center for Applied Photonics (CAP).
  
  \appendix
  
  \section{Projectors}\label{sec:projectors}
  
  Here we explain how to find a compact form for projection operators on spin-valley states which have a particular structure, such as the $(1,1)$ antisymmetric states described in this paper. For the sake of clarity we list them here explicitly:
  \begin{subequations}
    \label{eqn:antisymmstates}
    \begin{align}
      \label{eqn:fuau}
      \ket{\fu\au} & = (\ket{\ku\su;\ku\sd} - \ket{\ku\sd;\ku\su})/\sqrt{2}, \\
      \label{eqn:fdau}
      \ket{\fd\au} & = (\ket{\kd\su;\kd\sd} - \ket{\kd\sd;\kd\su})/\sqrt{2}, \\
      \label{eqn:aufu}
      \ket{\au\fu} & = (\ket{\ku\su;\kd\su} - \ket{\kd\su;\ku\su})/\sqrt{2}, \\
      \label{eqn:aufd}
      \ket{\au\fd} & = (\ket{\ku\sd;\kd\sd} - \ket{\kd\sd;\ku\sd})/\sqrt{2}, \\
      \label{eqn:nupplus}
      \ket{\nup_+} & = (\ket{\ku\su;\kd\sd} - \ket{\kd\sd;\ku\su})/\sqrt{2}, \\
      \label{eqn:nupminus}
      \ket{\nup_-} & = (\ket{\ku\sd;\kd\su} - \ket{\kd\su;\ku\sd})/\sqrt{2},
    \end{align}
  \end{subequations}
  with $\ket{\tau_1\sigma_1 ; \tau_2\sigma_2} = \cre{L\tau_1\sigma_1} \cre{R\tau_2\sigma_2} \ket{0}$. The operators that project on these states are shown in Eq.~\eqref{eqn:antisymmprojectors}. We can follow two approaches to obtain them, one is more intuitive and gives compact results, assembling projectors on larger parts of the Hilbert space, but in practice it works only when the states we are considering have a structure for which we already know the correct basic projectors. The other one is rather formal but general, however it does not give the projectors in a compact form.
  
  In order to show that equations \eqref{eqn:antisymmprojectors} are really the projectors we are looking for, we will present here the intuitive way to derive them, using symmetries in the structure of states \eqref{eqn:antisymmstates}. Take $\tau_z$, this operator has eigenvalue $+1$ when the state is in valley $\ket{\ku}$ and eigenvalue $-1$ when the state is in valley $\ket{\kd}$. Thus, $(\id + \tau_{jz}) / 2$ is the projector on all those states which have an electron in the $j$-th QD in valley $\ket{\ku}$ and $(\id - \tau_{jz}) / 2$ projects on states where the electron in QD $j$ has valley $\ket{\kd}$. Analogous considerations hold for the spin operator $\sigma_z$ and the spin states $\ket{\su}$, $\ket{\sd}$. In other words, we wrote down projectors on states which have a certain valley or possess a certain spin in a specific dot. Now focus on $\tau_{Lz}\tau_{Rz}$, this string of operators has eigenvalue $+1$ when both electrons are in the same valley ($\ket{\ku\ku}$ or $\ket{\kd\kd}$) and eigenvalue $-1$ when the valleys are different ($\ket{\ku\kd}$ or $\ket{\kd\ku}$). Then, operators $(\id \pm \tau_{Lz}\tau_{Rz}) / 2$ project on states whose valleys are the same ($+$) or are opposite ($-$). Similar considerations are valid for their spin counterparts.
  
  The states with the simpler structure are $\ket{\fu\au}$, $\ket{\fd\au}$, $\ket{\au\fu}$ and $\ket{\au\fd}$. They all have one of the properties fixed for both dots (either positive valley, negative valley, spin up or spin down, respectively). Consider $\ket{\fu\au}$ (Eq.~\eqref{eqn:fuau}), for both dots the valley is $\ket{\ku}$, but the spins are in a singlet state. The combination
  \begin{equation}
    \frac{1}{4} (\id + \tau_{Lz}) (\id + \tau_{Rz})
  \end{equation}
  projects on all the states with positive valley in both dots. To complete the expression we multiply by the projector on the spin singlet, given by $(\id - \vect{\sigma}_L \cdot \vect{\sigma}_R)/4$. Finally, the complete projector is
  \begin{equation}
    P_{\ket{\fu\au}} = 
    \begin{multlined}[t][.7\columnwidth]
      \frac{1}{16}
      \left ( \id + \tau_{Lz} \right )
      \left ( \id + \tau_{Rz} \right )
      \left ( \id - \vect{\sigma}_{L} \cdot \vect{\sigma}_{R} \right ),
    \end{multlined}
  \end{equation}
  as in \eqref{eqn:projfufdau}. The other very similar expressions in \eqref{eqn:projfufdau} and \eqref{eqn:projaufufd} follow the same derivation with the appropriate changes of signs and spin/valley operators.
  
  States $\ket{\nup_\pm}$ of equations \eqref{eqn:nupplus} and \eqref{eqn:nupminus} have a more complicated structure. First of all, they are both composed of states with opposite valley and opposite spin in the dots. The operator
  \begin{equation}
    \frac{1}{4} (\id - \tau_{Lz}\tau_{Rz}) (\id - \sigma_{Lz}\sigma_{Rz})
  \end{equation}
  projects on the subspace spanned by $\{\ket{\ku\su;\kd\sd}, \ket{\kd\sd;\ku\su}, \ket{\ku\sd;\kd\su}, \ket{\kd\su;\ku\sd}\}$. To distinguish between $\ket{\nup_+}$ and $\ket{\nup_-}$, we note that for $\ket{\nup_+}$ the product of the eigenvalues of spin and valley inside each dot is positive, while for $\ket{\nup_-}$ it is negative (this is why they get an energy shift of $\pm 2\Delta$ by the spin-orbit coupling $\hamspinsplit$). We use $\tau_{Lz}\sigma_{Lz}$ to distinguish them and the projectors become
  \begin{equation}
    \frac{1}{8} (\id - \tau_{Lz}\tau_{Rz}) (\id - \sigma_{Lz}\sigma_{Rz})
    (\id \pm \tau_{Lz}\sigma_{Lz}),
  \end{equation}
  positive sign for the subspace spanned by $\{\ket{\ku\su;\kd\sd}, \ket{\kd\sd;\ku\su}\}$ and negative sign for the subspace spanned by $\{\ket{\ku\sd;\kd\su}, \ket{\kd\su;\ku\sd}\}$. Finally, the operator that identifies the correct superposition is $\tau_{Lx} \sigma_{Lx} \tau_{Rx} \sigma_{Rx}$, the same for both states. The total projectors for $\ket{\nup_\pm}$ are
  \begin{equation}
    P_{\ket{\nup_\pm}} = 
    \begin{multlined}[t][.7\columnwidth]
      \frac{1}{16}
          \left ( \id - \tau_{Lz} \tau_{Rz} \right )
          \left ( \id - \sigma_{Lz} \sigma_{Rz} \right ) \\ \times
          \left ( \id \pm \tau_{Lz} \sigma_{Lz} \right )
          \left ( \id - \tau_{Lx} \sigma_{Lx} \tau_{Rx} \sigma_{Rx} \right ),
    \end{multlined}
  \end{equation}
  as in Eq.~\eqref{eqn:projnup}.
  
  Formally, the above results can also be obtained by noticing that the projectors we are looking for are spin-and-valley operators (i.e. acting on the $(1,1)$-subspace) and every spin-and-valley operator $P$ can be expanded in a linear combination of Pauli operator strings,
  \begin{equation}
    P = \sum_{i, j, k, l = 0, x, y, z} P_{ijkl} \;
    \tau_{Li} \sigma_{Lj} \tau_{Rk} \sigma_{Rl},
  \end{equation}
  where $P_{ijkl}$ are complex coefficients. If $\ket{\psi}$ is the state we want to project on, the projector is $P = \ket{\psi}\bra{\psi}$ and we can use the inner product provided by the trace operation to obtain the expansion coefficients,
  \begin{equation}
    P_{ijkl} = \mathrm{Tr} \left [ \ket{\psi}\bra{\psi} 
    \tau_{Li} \sigma_{Lj}  \tau_{Rk}  \sigma_{Rl} \right ].
  \end{equation}
  Again, recollecting a compact expression from this set of coefficients requires some work (even when most of them are zeros) and does not give any physical insight.
  
  For completeness, we briefly discuss $P_{\text{as}}$, the projector on the whole antisymmetric subspace of $(1,1)$-states of Eq.~\eqref{eqn:antisymmproj}. Although this operator could be obtained as the sum of the 6 projectors of Eq.~\eqref{eqn:antisymmprojectors} on the single, linearly independent antisymmetric states, we can write it down following another procedure, derived in Ref.~\cite{rohling_universal_2012}. It can be expressed as a combination of projectors on the singlet ($S$, antisymmetric) and on the triplet ($T$, symmetric) subspaces for spin and valley,
  \begin{equation}\label{eqn:asymmprojexplained}
    P_{\text{as}} = 
    P_{\text{spin}}^S P_{\text{valley}}^T + 
    P_{\text{spin}}^T P_{\text{valley}}^S.
  \end{equation}
  Here we defined the spin projectors as
  \begin{equation}
    P_{\text{spin}}^S = \frac{1 - \vect{\sigma}_L \cdot \vect{\sigma}_R}{4},
    \qquad
    P_{\text{spin}}^T = \frac{3 + \vect{\sigma}_L \cdot \vect{\sigma}_R}{4}.
  \end{equation}
  The valley projectors have the same form, only substituting $\vect{\tau}$'s for $\vect{\sigma}$'s. Eq.~\eqref{eqn:asymmprojexplained} yields, then,
  \begin{equation}
    P_{\text{as}} = (3 - \vect{\sigma}_L \cdot \vect{\sigma}_R - 
    \vect{\tau}_L \cdot \vect{\tau}_R - 
    (\vect{\sigma}_L \cdot \vect{\sigma}_R)(\vect{\tau}_L \cdot \vect{\tau}_R)) / 8.
  \end{equation}
  
  \section{Conditions on spin and valley Zeeman coupling constants for (1,1)-subspace}\label{sec:lesconditions}
  
  In Sec.~\ref{sec:spinvalleyzeemanoneone} we discussed the effects of the spin and valley Zeeman term on the exchange interaction in the case when the $(1,1)$-subspace is our LES. There we omitted to show the conditions to ensure that $(1,1)$-states are lower in energy than $(2,0)$ and $(0,2)$ states. We report them here. To simplify the discussion we assume that the spin Zeeman coupling constants associated with $x$- and $y$-Pauli matrices are small: $|h_{Sji}| \ll U$, $j = L, R$, $i = x, y$. Then, choosing an appropriate basis where $\sigma_{jz}$ and $\tau_{jz}$ are diagonal, it is easy to see that $(1,1)$-states are lower under the condition,
  \begin{multline}
    \max\{
    -2\Delta + |\HS{z} - \HV{z}|,
    -2\Delta + |\dhS{z} - \dhV{z}|, \\
    +2\Delta + |\HS{z} + \HV{z}|,
    +2\Delta + |\dhS{z} + \dhV{z}|,\\
    |\dhS{z} + \HV{z}|,
    |\dhS{z} - \HV{z}|, \\
    |\HS{z} + \dhV{z}|,
    |\HS{z} - \dhV{z}|
    \} + \\
    + 2\max\{|\Delta|, |h_{VLz}|, |h_{VRz}|, |h_{SLz}|, |h_{SRz}|\}
    < U - |\varepsilon|,
  \end{multline}
  where we used the same notation described in Eq.~\eqref{eqn:spinvalleynotation}.

  % \bibliography{bibliography}
  

\end{document}